# Macroeconomic and Financial Management in an Uncertain World: What Can We Learn from Complexity Science?

*Thitithep Sitthiyot\**
*Public Debt Management Office, Ministry of Finance, Thailand*

## Abstract

⒯his paper discusses serious drawbacks of existing knowledge in macroeconomics and finance in explaining and predicting economic and financial phenomena. Complexity science is proposed as an alternative approach to be used in order to better understand how economy and financial market work. This paper argues that understanding characteristics of complex system could greatly benefit financial analysts, financial regulators, as well as macroeconomic policy makers.

---
\* **Address:** Rama VI Rd. Phayathai, Bangkok 10400. Email: thitithep@mof.go.th

# การบริหารจัดการเศรษฐกิจมหภาคและการเงินในโลกของความไม่แน่นอน: เราสามารถเรียนรู้อะไรได้จากวิทยาศาสตร์ของความซับซ้อน

ฐิติเทพ สิทธิยศ

*สำนักงานบริหารหนี้สาธารณะ กระทรวงการคลัง ประเทศไทย*

## บทคัดย่อ

บทความนี้อธิบายถึงข้อบกพร่องที่สำคัญขององค์ความรู้ทางเศรษฐศาสตร์มหภาคและการเงินที่ใช้ในการอธิบายและทำนายปรากฏการณ์ทางเศรษฐกิจและการเงิน พร้อมกับนำเสนอและแสดงให้เห็นว่า วิทยาศาสตร์ของความซับซ้อนเป็นทางเลือกสำหรับใช้ศึกษาและเข้าใจการทำงานของระบบเศรษฐกิจและการเงินได้ดีกว่า บทความนี้ตั้งข้อสังเกตว่าการเข้าใจคุณสมบัติของระบบที่มีความซับซ้อนจะเป็นประโยชน์อย่างยิ่งต่อนักวิเคราะห์ทางการเงิน หน่วยงานกำกับดูแลทางการเงิน และผู้กำหนดนโยบายเศรษฐกิจมหภาค

"We cannot solve our problems with the same kind of thinking that created them."
- Albert Einstein

**คำสำคัญ:** เศรษฐศาสตร์มหภาค; การเงิน; วิทยาศาสตร์ของความซับซ้อน

# 1. Introduction

Macroeconomics and finance have been widely criticized by a number of scientific scholars as well as economists for the past several decades regarding drawbacks and limitations of models and assumptions used in order to analyze, explain, as well as predict macroeconomic and financial phenomena.[1] However, those criticisms have hardly affected the ways in which orthodox economists and financial analysts use to analyze macroeconomic and financial issues.[2] Worse, if one asks orthodox economists and financial analysts about shortcomings and limitations of models and assumptions used in macroeconomics and finance, one might be surprised to find out that many of them have not even come across such criticisms. Or if they have, those orthodox economists and financial analysts still insist that their models and assumptions are useful framework and keep on using them despite the fact that these models have repeatedly failed to explain and predict macroeconomic and financial events.

Ironically, this trend seems to go on unnoticed by policy makers who normally take advices from orthodox economists and financial analysts. It is as if an engineer had used a model along with its assumptions in designing and constructing ten bridges and all bridges built in the past collapsed and the same engineer is still being trusted and commissioned to design and build the eleventh bridge! In sciences, if models and assumptions have flaws in a sense that they cannot explain and predict empirical observations, good scientists will search for better models and/or change assumptions until they find models and assumptions that could explain and predict real world phenomena. Unfortunately, this is not the case for orthodox macroeconomics and modern finance since there has not been any significant changes since the U.S. financial crisis in 2007. The knowledge is still being taught in the universities and being used in formulating economic and financial policies. This paper tries to argue that we should abandon them due to their serious drawbacks.[3] It then discusses and raises the importance of an alternative approach that could help scientifically analyze, explain, and predict macroeconomic and financial phenomena.

This paper is organized into five sections. Following the introduction, Section II discusses five selected but critical shortcomings of existing knowledge in macroeconomics and finance. Despite a number of well-known scientists as well as many economists have

---

already pointed out in many occasions as mentioned above, this paper believes it is worth to reiterate the importance of this issue. The drawbacks of knowledge in macroeconomics and finance would help lay the ground for Section III where the paper introduces the concept of science of complexity as an alternative approach and describes how it could help us understand the nature of financial market and economic system. The implications of complexity science for macroeconomic and financial uncertainty management are discussed in Section IV. Finally, Section V concludes and suggests future direction for macroeconomics and finance.

## 2. Drawbacks in Macroeconomics and Finance

The main shortcomings in macroeconomics and finance have to do with models and assumptions. For macroeconomics, the model that is being widely used by economists to analyze and forecast the effects of economic policies on the economy is known as the Dynamic Stochastic General Equilibrium or DSGE model.[4] This model has been heavily criticized by many scientists outside the field of economics and by a number of economists that many assumptions imposed in the DSGE model are not consistent with empirical observations.[5] This paper chooses to discuss three main assumptions imposed in the DSGE model that do not fit with what happen in reality, namely, the equilibrium state of the economy, the external random shocks as the only factor that could affect the system, and the representative agent with rational expectations.

As its name suggests, the DSGE model assumes an equilibrium state of the economy. Casti (2010) illustrates that, in the real world, an economy where supply and demand are in balance never happens even approximately. According to Helbing (2015), economic system is unlikely to be in equilibrium at any point in time. Rather, it is expected to show a complex non-equilibrium dynamics. Ball (2012) notes that the equilibrium assumption originates from microeconomic theory as an analogue of equilibrium physical systems such as gases which have stable and unchanging states. The physical sciences, however, have long moved on to describe non-equilibrium process such as weather system but economics has not. According to Ball, the equilibrium assumption is one of the key reasons why the DSGE model not only failed to forecast major fluctuations such as slumps and crashes but also motivated the disastrous suggestions by many politicians and economists before the 2007 U.S. financial crisis that such crises had been banished for good.[6] In addition, Arthur (2015) points out that, by assuming an equilibrium state of the economy, economists place a very strong filter on what we can see in the economy. Under equilibrium, by definition, there is no scope for improvement or further adjustment, no scope for exploration, no scope for creation, and no scope for transitory phenomena. Thus, anything in the economy that takes adjustment, namely, adaptation, innovation, structural change, and history itself, must be bypassed or dropped from the theory. The result may be a beautiful structure but it lacks authenticity, aliveness, and creation. It is no wonder why Charles Goodhart, the former Monetary Policy Committee of the Bank of England, articulates that "the DSGE model only works when nothing happens [emphasis added]."[7] If we look around the world in which we live, we do observe things continue to unfold and hardly stay in equilibrium as assumed in the DSGE model.[8]

---

[4] Please see Tanboon (2008) and Sbordone et al. (2010) for general description of the DSGE model.
[5] For example, please see Ormerod (2006), Beinhocker (2007), Colander et al. (2008), Buiter (2009), Farmer and Geanakoplos (2009), Krugman (2009), Caballero (2010), Düppe (2010), Kirman (2010), Orrell (2010), Farmer (2011), Ball (2012), Ormerod (2012), Arthur (2013), Buchanan (2013), Goodhart, Tsomocos, and Shubik (2013), Arthur (2015) and Helbing (2015). Ormerod (2006) refers to Kenneth Arrow, the winner of the Sveriges Riksbank Prize in Economic Sciences in Memory of Alfred Nobel in 1972, who regards the DSGE model as being empirically refuted. Düppe (2010) interprets the banquet speech given by Gerard Debreu in 1983 upon receiving the Sveriges Riksbank Prize in Economic Sciences in Memory of Alfred Nobel as acknowledging the irrelevance of the general equilibrium theory and real world economic content.

While the DSGE model assumes equilibrium state of the economy, it does allow the economic system to be out of equilibrium. However, anything that forces the economy away from equilibrium, according to this model, must come from outside of the system only. In the DSGE model, when the economy is perturbed by an external shock, it would adjust towards a new stable equilibrium.[9] In the absence of an external shock, supply and demand would always be in balance, all markets would clear, and prices would be stable (Ball, 2012). This model is not capable of generating shock from the inside that causes the system to fluctuate.

Kirman (2010) argues that, by and large, the fluctuations of the economy are the result of interaction among agents who make up the system and not due to some exogenous shocks. Kirman also refers to Sornette (2003) who makes a similar point that a stock market crash is not the result of short-term exogenous events but rather involves a long-term endogenous build-up with exogenous events acting merely as triggers. The idea that endogenous factors could gradually cause the system out of equilibrium is not entirely new, however. It has long been recognized and studied by many disciplines such as physics, biology, ecology, and sociology. Furthermore, it should be noted that, in such systems, abrupt changes are not unusual and could be the result of spontaneous emergence of extreme events in self-organizing systems (Sornette, 2003). At the time when change occurs, normally with a long period of delay, the system may switch rapidly from one state to another, the so-called phase transition, and this would be dependent on endogenous factors comprising the system and not necessarily on some major exogenous shocks. By assuming an outside force as the only factor that could push the economy away from equilibrium, the DSGE model therefore does not allow one to analyze the dynamic process of the interaction among endogenous factors that could gradually self-organize, cause such a change as phase transition, and push the economic system out of equilibrium.

---

[6] Ben Bernanke, the former Chairman of the Federal Reserve Board, states in his remarks at the meetings of the Eastern Economic Association in Washington, D.C. on February 20, 2004 that the economic landscape over the past twenty years has faced stability and "great moderation."

[7] Personal comment given at the Bank of Thailand International Symposium, Bangkok, October 16, 2010.

[8] This is consistent with the second law of thermodynamics which states that systems tend to move from order to disorder as measured by entropy. Beinhocker (2007) argues that, given the fact that economic system exists in the real physical world, therefore, it must obey the second law of thermodynamics as everything else in the universe does. Beinhocker also refers to Sir Arthur Eddington, the British Astrophysicist, who remarks that "if your theory is found to be against the second law of Thermodynamics I can give you no hope; there is nothing for it but to collapse in deepest humiliation."

[9] Helbing and Balietti (2010) argue that, in a dynamic system of highly non-linearly interacting and coupled variables like an economy, there is no guarantee that the system would converge to a new stable

In addition, the DSGE model assumes that the whole population in the economy can be represented by a representative agent with rational expectations who tries to maximize expected utility at any given period subject to inter-temporal budget constraint. It is as if one person's thought can be used to represent the way in which everyone else in the entire economy thinks. Clearly, the model ignores the interaction among different agents comprising the economy and positive feedback which could cause emergent phenomena.

By looking superficially, the assumption about human rationality might seem to make sense. However, any sane person might be shocked into disbelief if he/she finds out about the thought process of the representative agent with rational expectations before making a decision. This is very interesting because, as of today, many scholars still employ this assumption as a building block for their models whether it is a DSGE type or not.[10] Beinhocker (2007) provides a very vivid example about the thought process of a representative agent with rational expectations before making any decisions. Imagine a representative agent in the DSGE model walks into a local grocery store and sees some tomatoes. If you are a representative agent with rational expectations as the DSGE model assumes, the followings are your logical thought process while deciding whether or not to buy tomatoes.

First, you have a well-defined preference for tomatoes compared with everything else in the world you could possibly buy. Beinhocker uses bread, milk, and a vacation in Spain as examples. In addition, you have well-defined preferences for everything you could buy at any point in the future. Since the future is uncertain, you have to assign probabilities to those potential purchases. For instance, you might anticipate that there is a 23% chance that, in the next two years, the shelf in your kitchen will come loose and you will need to pay US$ 1.20 to buy some bolts to fix it. The discounted present value of US$ 1.20 is about US$ 1.00, multiplied by a 23% probability, equals an expected

---

equilibrium. It is also possible that the system would have multi-equilibria. In addition, the existence of equilibrium does not necessarily imply that it is stable where the system will converge to this solution. The stable solution could be a focal point with orbiting solutions as explained by Lotka-Volterra equations or it could be unstable and give rise to a limit cycle or a chaotic solution.

value of 23 cents for possible future repairs, which you must trade off with your potential purchase of tomatoes today, along with all of your other potential purchases in your lifetime. In the DSGE model, all your well-defined preferences are also ordered very logically. For example, if you prefer tomatoes to carrots, and prefer carrots to green beans, you will always take tomatoes over green beans.

In the DSGE model, it assumes that you know exactly what your budget constraint is for spending on tomatoes. To calculate this budget, you must have fully formed expectations of your future earnings over your entire lifetime and have optimized your current budget based on that knowledge. You might not buy those tomatoes because you know that the money spent on them could be better spent in your retirement. The model also assumes that your future earnings will be invested in a perfectly hedged portfolio of financial assets and that you already take into account actuarial calculations on the probability that you will live until retirement at the age of sixty-five, as well as your expectations of future interest rates, inflation, and the yen-to-dollar exchange rate.

While you are standing in the grocery store, looking at those nice, red tomatoes, you then feed all these information into your mind and perform an incredibly complex optimization calculation that trades off all these factors, and come up with the perfectly optimal answer - to buy or not to buy tomatoes!

It is quite obvious that no one would do this kind of analysis while deciding whether or not to buy goods and services. According to Simon (1983), no human being would be able to perform such a complex calculation. It is being treated by traditional economics as if human beings perform such complex task while, in reality, no one does. Even if there were one[11], there is still a limitation in the sense that this kind of analysis would require an infinite amount of information in order to perform such calculation. In the real world, the cognitive capacities of human are bounded and abilities to memorize facts and to perform complicated logical analyses are very limited as argued by Helbing and Balietti (2010). Helbing and Balietti also note that this is a non-deterministic polynomial-time hard problem (N-P hard problem) even computers are facing limits where optimization cannot be performed. Axelrod (1997) has long noted that the reason for the dominance of assumption about human rationality is not that scholars think it is realistic. Rather, it allows for deduction. Axelrod views that this unrealistic assumption undermines much of its value as a basis for advice. Besides Axelrod (1997), Simon (1983) and Helbing and Balietti (2010), a number of psychologists, neuroscientists, and behavioral economists, as well as scholars and practitioners from various disciplines have argued in similar fashion.[12]

---

[11] From the viewpoint of Buchanan (2013), that person might be insane.

[12] For example, please see Kahneman and Tversky (1981), Thaler (1994), Ariely (2009), Bardsley et al. (2010), McClure (2011), Santos (2011), McFadden (2013), Sutherland (2014), Fiske (2015), Haidt (2015), Knutson (2015), Levi (2015), Pentland (2015), Slingerland (2015), and The World Bank (2015) for criticisms about human rationality as assumed in traditional economics.

The critical point is that if the assumptions about the equilibrium state of the economy, the external shock as the only factor that causes the economic system out of equilibrium, and the representative agent with rational expectations as imposed in the DSGE model are not close to what happen in reality, then any estimated impacts of policies on the economy and/or economic forecasts coming out of this model, even if they are correct, must be a coincidence, not because of skills of the analysts based on their scientific knowledge of reality. In fact, if one compares economic forecasts using the DSGE model and actual numbers, one would find that they are not very close, and most of the time when that happens, external factors always take the blame, not skills of the forecasters.[13] The Economist points out that the DSGE model was a poor guide to the origin of the financial crisis and left its followers unprepared for the symptoms (The Economist, 2009). Kirman (2010) also argues that if the DSGE proponents think that their models are useful for understanding how economy works, then they should be able to explain why their models do not allow for the possibility of a crisis. If major crises are a recurrent feature of the economy as documented by Reinhart and Rogoff (2009), then the DSGE model should incorporate this possibility.

Based on the reasons discussed above, this paper argues that if macroeconomics wants to progress, the DSGE model along with its assumptions about the equilibrium state of the economy, the exogenous shock as the only force that causes the economic system out of equilibrium, as well as the representative agent with rational expectations have to be abandoned and replaced with a better model and more realistic assumptions.

While the drawbacks of macroeconomics are due mainly to the model and assumptions, the problem in finance has to do with assumptions. There are two assumptions this paper chooses to discuss. The first assumption often imposed in finance is that price changes are independent. One could think of tossing a fair coin as a metaphor. The results coming out of a coin tossing are independent from each other. The coin does not have a memory whether it landed head or tail in the past. The second assumption is about the distribution of data. Most financial models and tools assume that data are normally distributed.[14] Ho Ho pointed out that we can only find these two statistical properties of financial data in our dreams.[15] Fernandez (2015) also views these two assumptions as absurd. In reality, changes in prices are not independent. Financial data have a property of path dependence or long memory. Normally, big changes are followed by big changes, the so-called clustered volatility[16], but it is very difficult to predict whether it is going up or down. Benoit Mandelbrot notes that if each change in price, whether a five-cent uptick or a US$ 26 collapse, appears independently from the last, and price changes last week or last year do not influence those today, it would mean any information that could be used to predict tomorrow's price is contained in today's price. In that case, there is no need to study historical charts.[17]

---

[13] Ormerod (2012) criticizes that despite apparent intellectual advances of the DSGE model, forecasters continue to make the same mistakes since 1970s.

[14] Capital asset pricing model, efficient market hypothesis, and Black-Scholes option pricing formula are prime examples of such models and tools.

[15] Comment given at the 4th Annual National Asset-Liability Management Conference, Singapore, July 7, 2011.

[16] According to Farmer and Geanakoplos (2009), clustered volatility is referred to the fact that there are substantial and strongly temporally correlated changes in the size of price movements at different point in time.

In addition to the assumption that price changes are independent which is not consistent with empirical observations, financial data do not follow normal distribution as assumed in modern finance. Rather, it exhibits power law distribution with fat tails. According to Haldane and Nelson (2012), the power law distribution with fat tails implies that the probability of large events decreases polynomially with their size while, in the normal distribution world, the probability of large events declines exponentially with their size, making large events increasingly rare at a rapid rate. In contrast, under power law distribution, these large events are much more likely.

To provide a numerical example of how risky it might be if one assumes normality of distribution of data, this paper refers to a study conducted by Benoit Mandelbrot using daily index movement of Dow Jones Industrial Average during the period of 1916-2003.[18] Based on Mandelbrot's empirical findings, normal distribution implies that there should be fifty-eight days when the Dow moves more than 3.4% while in fact there are one thousand and one. In addition, normal distribution predicts six days where the Index swings beyond 4.5% whereas there are three hundred and sixty-six days according to the empirical observations. And lastly, the Index that swings more than 7% should come once every three hundred thousand years as predicted by normal distribution while the twentieth century already observed forty-eight days. It is crystal clear based on this empirical evidence that assuming that data have normal distribution would highly underestimate risk. It is so worried that, nowadays, many people both in academic and in practice have yet to realize about these limitations and continue to teach or do their business as if nothing has happened over the past eight years.

This paper views that these are serious issues that cannot be taken for granted. The failures in macroeconomics and finance, according to Helbing and Balietti (2010), are not a matter of approximations which often lead to the right understanding but wrong numbers. Rather, it concerns fundamental errors in the sense that certain conclusions following from these models along with their unrealistic assumptions are seriously misleading. As the recent financial crisis in the U.S. has demonstrated, such errors in macroeconomics and finance can be extremely costly, US$ 4-20 trillion based on the number estimated by Helbing (2009).

How should the disciplines of macroeconomics and finance be changed in order to avoid repeating the same mistakes in the future? This leads to the third topic of this paper where it introduces the concept of complexity science. This paper argues that this new approach has strong potential and should be the direction that macroeconomics and finance should move towards.

---

[17] Mandelbrot and Hudson (2008).
[18] Ibid.

## 3. Complexity Science as an Alternative Approach to Understand the Workings of Economy and Financial Market

Complexity science is a branch of science that studies properties of complex systems. Examples of complex systems are earthquake, sand pile avalanche, forest fire, weather system, ecosystem, food web, epidemics, traffic flows, neuron network, social network, modern political uprising, financial market, and socio-economic system. If one observes these systems, one would find that they are characterized by interactions of elements comprising the systems. These interactions, either directly or indirectly, are highly non-linear and tend to be dynamic and probabilistic with positive feedback[19] which often generates surprises or emergent phenomena that a single element in the system would not be able to do or create it on its own[20] (Holland, 1996). Mitchell (2009) defines complex system as a system in which large networks of components with no central control and simple rules of operation can give rise to complex collective behaviour, sophisticated information processing, and adaptation via learning or evolution. According to Mitchell, in such a system where organized behaviour arises from the bottom up without an internal or external controller or central planner are called self-organizing.[21] Since simple rules produce complex behaviour in hard-to-predict ways, the macroscopic behaviour of such system is called emergent.

Given characteristics of complex system as described above, it is very difficult or impossible to predict what is going to happen in the complex system, making it difficult to control and prevent the event from happening. In addition, it is not very easy to find the true cause of the event in the complex system. For example, in nature, when an avalanche occurs, one cannot blame a single snowflake. Likewise, an extremely volatile movement in a stock market, the so-called flash crash in the modern era, could happen for no special reason.[22] Moreover, financial and economic crises could arise locally from the interactions of agents making up the system but none of them can really be held responsible for such crises as argued by Kirman (2010).

---

[20] This implies that the whole is not equal to the sum of its parts.

[21] It should be noted that, in general, complex system needs no designed but, nonetheless, can be designed. Urban planning and development is one example of designed complex systems.

[22] This could be illustrated by closed experiment based on the El Farol Problem and the Minority Game where the winners in the experiment are the minority. The author of this paper has conducted a number of closed experiments over the years and the results are always qualitatively similar in the sense that the ratio of persons who are in minority group always fluctuates, sometimes excessively, without any presence of external factors or special reasons. For further readings regarding the El Farol Problem and the Minority Game, please see Arthur (1994), Challet and Zhang (1997), Ball (2005), Beinhocker (2007), and Buchanan (2013).

Another interesting thing that we can learn from the characteristic of complex system is that while the event is happening, it is very difficult to tell how and when it is going to end. Furthermore, assessing the impact of the event while it is not fully finished could give a wrong picture about the size of the impact since it could be quite different. The losses from tsunami in Japan and the damages from flood in Thailand both in 2011 should provide good examples of characteristics that complex systems share in common. The same is true for financial and economic crises. We have yet to know how and when the problems in the U.S. economy and in some European countries are going to end and probably lost count the magnitude of losses resulting from the 2007 U.S. financial crisis.

It should also be noted that complex system exhibits path dependence where history matters. Path dependence along with highly non-linear relationship among elements comprising the system would make the system very sensitive to initial state. A small change in the initial state could lead to totally different outcome through positive feedback. This is known as the butterfly effect coined by a meteorologist named Edward Lorenz. A flap of butterfly's wings in Brazil could cause a tornado in Texas. Likewise, a bankruptcy of one financial institution could cause the world financial system to collapse, often times without any warnings.

According to Haldane and Nelson (2012), complex system could organize itself to a fragile state. Bak (1996) calls this state self-organized criticality and uses sand pile avalanche to explain this phenomenon. Imagine dropping a grain of sand one by one on the flat surface. As more grains are added, the pile builds up. At some point, adding another grain would cause the pile to collapse. Sand pile reaches its self-organized critical state. Haldane and Nelson (2012) argue that the build-up to the 2007 U.S. financial crisis has the same characteristic as that of sand pile avalanche.

A competitive search for yield among financial firms caused them to increase risk to hit return targets. The sand pile of returns grew ever-higher as the system self-organized on a high-risk state. Having reached this critical state, a small addition of risk - a single subprime grain of sand - was sufficient to cause the whole edifice to collapse in an uncontrolled cascade.

Haldane and Nelson (2012) also note that the size of sand pile avalanches and the magnitude of losses due to financial crises, like most things happening in natural, social, economic, and financial systems, namely, the monthly total rainfall, the intensity of solar flares, the magnitude of earthquakes, the number of citations, the frequency of occurrence of words in a book, the number of population in cities, income distribution, as well as annual real bank loan growth and real GDP growth among different countries, are observed to have power law distributions with fat tails.

Mathematically speaking, power law distribution of a given variable describes an inverse relationship between the size of that variable and its frequency.[24] The existence of power law distributions among variables in natural, social, economic, and financial systems indicates that small events are more common than large ones. However, the interesting characteristic of power law distribution is that it is scale-independence which means that the events of all sizes can be generated by the same process. For example, the biggest and most infrequent earthquakes are created by the same processes that produce hordes of tiny ones. Likewise, the occasional stock market crashes are generated by the same internal dynamics of the marketplace that produce daily movements in stock prices (Ball, 2012).[25] This makes it hard to predict how small or large any particular event will become. All we can say is that large event is less frequent than small one, nothing more.

Given characteristics of power law distribution as described above, it implies that one cannot use mean as an estimate for the size of the future event and cannot use variance or standard deviation to measure the risk associated with that event since these estimates are not well defined.[26] Thus, the average size of avalanche is not helpful in predicting the size of the next avalanche. Neither is the variance useful as a measure of the size distribution of the avalanches. This is similar to what happen in financial and economic crises where average loss of previous crises is not a good predictor for the loss of the next crisis. Nor is variance useful for macroeconomic and financial risk management. Alan Greenspan notes that each crisis is different and none of them will look alike.[27]

By viewing financial and economic systems from the lens of complexity science, it is clear that this alternative approach provides far superior descriptions of how the financial market and the economy work than those postulated in the DSGE model and in modern finance theories. Flake (1998) argues that, considering an economist who builds a model of the economy on a computer and reaches the conclusion that interest rates, unemployment, inflation, and GDP growth will all reach a constant level at the end of the year and stay that way forever, it is obvious that this model tells us very little about how the real world works because the model fails to capture an important aspect of the real world. Instead, supposing a model is built in such a way that it never reaches equilibrium, turns out to be extremely sensitive to the starting conditions, and displays surprisingly complex behaviour, even if this model fails to make actual predictions about the real economy, it still has some predictive power since it may reveal a deeper truth about the inherent difficulty of predicting the economy.

---

[25] For further reading about the butterfly effect, please see Lorenz (1995).

[24] The mathematical form of power law distribution is $P(X > x) \sim x^{\alpha}$. That is the probability of a random variable $X$ exceeds some level $x$ is proportional to $1/x^{\alpha}$, suggesting that the probability of large events decays polynomially with their size.

[25] These two examples suggest that they are not outliers or one-off events.

[26] One way to detect whether or not mean and variance have any useful meanings is by adding a single observation into the sample. If that single observation completely alters the estimates of mean and variance of the sample, then mean and variance are not very useful.

[27] Comment given during the interview for BBC 2.

Building such a dynamic non-equilibrium model of economic system or any other complex system and then testing the conclusions of the model against real world observations are no longer beyond our reach. Given the advances of computing power of computer and of knowledge in computer science for the past few decades, scholars and practitioners from various disciplines are now able to analyze, explain, and predict characteristics and behaviours of complex systems by using computer simulation models. Examples of well-known computer simulation models are agent-based or individual-based models.[28]

According to Railsback and Grimm (2012), agent-based or individual-based models are models where individuals or agents are described as unique and autonomous that usually interact with each other and their environment locally. Agents may be organisms, humans, businesses, institutions, and any other entity that pursue a certain goal. Being unique implies that agents usually are different from each other in such characteristics as size, location, resource reserves, and history. Interacting locally means that agents usually do not interact with all other agents but only with their neighbours in geographic space or in some other kind of space such as a network. Being autonomous implies that agents act independently of each other and pursue their own objectives. Organisms strive to survive and reproduce. Traders in the stock market try to make money. Businesses have goals such as meeting profit targets and staying in business. Regulatory authorities want to enforce laws and provide public well-being. Agents therefore use adaptive behaviour and adjust their behaviour to current states of themselves, of other agents, and of their environment. The main advantage of the agent-based model is that it can address problems concerning emergence that arises from individual agents, each with simple rules, interacting and responding to each other and to their environment and that results from bottom-up and self-organized processes, not top-down directions.[29]

Given characteristics of complex system and models to analyze its properties using computer simulation, how could financial regulators, financial analysts, and macroeconomic policy makers benefit from understanding the nature of complex systems? Section IV discusses this issue.

---

[28] For further readings about agent-based modeling, please see Axelrod (1997), Flake (1998), Helbing (2012), and Railsback and Grimm (2012).

[29] Helbing (2009) notes that since complex systems cannot be controlled in conventional way like pressing a button or steering a car, top-down control attempts will usually fail. The right approach to influence behaviours in complex systems is to come up with rules that support and strengthen the self-organization and self-control of the systems so that coordination and cooperation in complex systems will appear by themselves without specifying what exactly each element in the systems should do.

## 4. Implications of Complexity Science for Macroeconomic and Financial Uncertainty Management

First of all, since complex systems are characterized by highly non-linear interactions among elements comprising the systems, they normally follow power law distribution with fat tails. Understanding this characteristic of complex system should help financial analysts, financial regulators, and policy makers be more careful when pricing financial products and/or conducting risk analysis. If done well, it could help lower the chance of financial and economic crises in the future. It is important to note that this does not mean that crisis will not happen again but the chance of happening could be reduced.

Another implication from complexity science is that an excessively volatile financial market could happen for no special reason. Simply an unnoticeable internal interactions among participants in the market could result in this phenomenon via a butterfly effect. For those who rely on fundamental analysis, understanding this characteristic of complex financial market would save time searching for external factors that might cause volatility in the market.

In addition, given the characteristic of emergence in complex financial and economic systems, regulators and policy makers must understand that, instead of spending time and effort in designing measures to prevent financial and economic crises, they should focus on mitigating measures since these crises cannot be prevented in the first place. Mandelbrot notes that after each financial and/or economic crisis, authorities always try to come up with new or adjusted rules and regulations in hope that these new or adjusted rules and regulations could help prevent the next crisis without realizing that this cannot be done.[30]

Rather than imposing rules and regulations with an aim to prevent the next crisis, this paper suggests that financial regulators and policy makers should explore how the immune system detects and kills pathogens without harming the body and apply that knowledge in designing rules and regulations to mitigate and/or eliminate problems after the financial market or economic system is being hit by the crisis. Thus far, a number of scholars and practitioners have begun to design economic and financial systems based on the insights from biology, ecology, as well as natural and life sciences. For example, the Bank of England, in cooperation with scientists, has started to draw comparisons between biological systems, with their complicated webs of interactions between all the different species, with the interactions between different banks and financial institutions (Durrani, 2011). Haldane and May (2011) explore the interplay between complexity and stability in financial networks by drawing analogies from the dynamics of ecological food webs and with networks within which infectious diseases spread. Helbing (2015) recommends that new economic system thinking should develop non-equilibrium network models that capture self-organized dynamics of real economic system. In addition, it should be an interdisciplinary approach that takes into account complex, ecological, and social system thinkings.

---

[30] Ibid.

Furthermore, regulators and policy makers must understand that, in complex systems, rare event is not unusual but norm.[31] Understanding this could help them prepare for the rare event by taking it into account when conducting risk analysis. Taleb (2013) suggests that this could be done by not considering the worst case scenario in the past but realizing that even worse outcome is possible and then adopting a redundancy strategy. Examples of redundancy strategy are maintaining extra reserves or savings, pursuing budget surplus, purchasing some forms of insurance, and hedging using financial derivatives. This is very critical for liability management. Taleb (2012) argues that if one has debt, one needs to be very accurate in his/her prediction about the future. By adopting redundancy strategy, it should immune the person from prediction errors.[32] The same can be applied for financial regulators and policy makers. For example, if the government runs a budget surplus and has accumulated a lot of savings, there is no need to predict the cause of the next crisis. This is because if a crisis does happen, it would be assured that the financial market and the economy are protected to some extent. Although they might be hit, the financial market and the economy should be in a relatively better position than those that have debts and are managed by people who do not understand characteristics of complex financial market and economic system and how they function. Unfortunately, this is not an easy task because financial market and economic system as we just learnt from complexity science that they are unpredictable and difficult-to-control even if we know how they work. The current financial crisis clearly demonstrates that we still do not know the workings of financial market and economic system very well.

Given the unpredictable and difficult-to-control nature of financial and economic systems, risk managers both in private and in public sectors should be prepared to deal with unexpected events that complex financial market and economic system will bring forth. What risk managers could do in a complex and adaptive financial and economic ecology is to check the environment very often and try to adjust strategies accordingly. More importantly, they must understand that they should not try to predict or determine behavior of a complex adaptive system, but to expect the most probable possibilities. This will make it easier for them to adapt when things go off the course. Because then, they are ready to expect the unexpected (Gershenson and Heylighen, 2004).

This paper views that the task of risk managers can be metaphorically compared to the task of performer walking on a high-wire across canyon. The performer has to constantly check the surrounding environment and tries to adjust and balance him/herself in order to avoid falling to the ground. The key difference is that, in the world of high-wire walking, when things go badly, performers could be seriously injured or lost their lives but, in the financial world, risk managers get massive bonuses at the expense of tax payers

---

[31] This is due to the characteristic of power law distribution of variables in complex systems.

[32] It is worth to note that redundancy is an important characteristic of nature that human beings should learn to emulate. We could survive with one kidney but nature gives us two. Some plants produce millions of pollen grains but most of them end up in our noses. A frog lays thousands of eggs but not all of them hatch into tadpoles, let alone, metamorphose into adult frogs. By creating debt, human being is the only known species so far whose behaviour is not consistent with that of all other natural living organisms.

## 5. Conclusions and Suggestions for Future Direction of Macroeconomics and Finance

In sum, the failures of existing knowledge in macroeconomics and finance in explaining and predicting economic and financial phenomena are mainly due to models and unrealistic assumptions these disciplines employ. This paper argues that the assumptions regarding equilibrium state of the economy, the exogenous shocks as the only factor that can push the economic system out of equilibrium, the representative agent with rational expectations as widely used in macroeconomics and the assumptions about independent changes in variables and normal distribution usually imposed in finance have (wrongly) guided economists, financial analysts, as well as policy makers into the illusions that the economy and financial market could be directed and controlled, and hence, made them unaware and unprepared for the abrupt changes.

In order for macroeconomics and finance to make progress, these knowledge have to be abandoned and replaced by other scientific methods and tools that could deal with complexity of macro-economy and financial market. This paper introduces complexity science as an alternative approach as well as the agent-based model which can be used to analyze and understand how economy and financial market function. It also discusses key important characteristics of complex system and the workings of agent-based model that economists, financial analysts, financial regulators, and policy makers must realize and include in their analyses. If done well, this paper believes that it would not only advance our understanding about the workings of economy and financial market but also reduce the chance of making mistakes that could cause economic and financial crises.

In addition, according to the United Nations (2013), the financial system has to be redirected towards promoting access to long-term financing for investments required to achieve sustainable development. This paper views the issue of sustainable development as one of the major challenges for regulators and policy makers in the years to come.[33] This paper agrees with Foxton et al. (2013) that, to achieve the goal of sustainable financial and socio-economic systems, regulators and policy makers must recognize that the existing knowledge in macroeconomics and finance as argued in this paper are not just irrelevant but positively harmful when the need is for systemic shifts in financial and socio-economic trajectories, and employ the knowledge from complexity science for their analyses.

---

[33] The author would like to thank Suradit Holasut for suggesting this metaphor.

However, we should realize that, in sciences, any significant changes in knowledge would take time simply because it is very difficult or almost impossible to convince those who believe that the earth is the center of the universe to see otherwise. But sooner or later this has to be changed. It is better to be sooner rather than later.

## Acknowledgement

The views expressed in this paper are of the author. They in no way reflect those of the Public Debt Management Office and/or Ministry of Finance. This paper has been evolved in an anti-fragile matter from the exclusive address entitled "Macroeconomic and Financial Uncertainty Management: What Can We Learn from Complexity Science?" given at the 6th National Asset-Liability Management Conference held in Singapore on July 24, 2013. The author would like to thank Suradit Holasut and Thomas D. Willett for their helpful comments. All errors are of the author.